\documentclass[epsf,12pt,russian]{article}

\usepackage[cp1251]{inputenc}
\usepackage[T2A]{fontenc}

\usepackage{amssymb}

\usepackage[dvips]{graphicx}

\begin{document}

\title
{Contribution of matter fields to the Gell-Mann-Low function for
N=1 supersymmetric Yang-Mills theory, regularized by higher
covariant derivatives.}

\author{K.V.Stepanyantz}

\maketitle

\begin{center}
{\em Moscow State University, physical faculty,\\
department of theoretical physics.\\
$119992$, Moscow, Russia}
\end{center}

\begin{abstract}
Contribution of matter fields to the Gell-Mann-Low function for
N=1 supersymmetric Yang-Mills theory, regularized by higher
covariant derivatives, is obtained using Schwinger-Dyson equations
and Slavnov-Tailor identities. A possible deviation of the
result from the corresponding contribution in the exact Novikov,
Shifman, Vainshtein and Zakharov $\beta$-function is discussed.
\end{abstract}


\section{Introduction.}
\hspace{\parindent}

It is well known, that supersymmetry essentially improves the
ultraviolet behavior of a theory. For example, even in theories
with unextended supersymmetry, it is possible to propose the form
of the $\beta$-function exactly to all orders of the perturbation
theory. This proposal was first made in Ref. \cite{NSVZ_Instanton}
as a result of investigating instanton contributions structure.
For the $N=1$ supersymmetric Yang-Mills theory with matter fields,
this $\beta$-function, called the exact
Novikov-Shifman-Vainshtein-Zakharov (NSVZ) $\beta$-function) is

\begin{equation}\label{NSVZ_Beta}
\beta(\alpha) = - \frac{\alpha^2\Big[3 C_2 - 2
C(R)\Big(1-\gamma(\alpha)\Big)\Big]}{2\pi(1- C_2\alpha/2\pi)},
\end{equation}

\noindent where $\gamma(\alpha)$ is the anomalous dimension of the
matter superfield. We note that here we present the result for a
theory with the Dirac fermions, and two chiral superfields
correspond to each of them. $R$ denotes a representation for one
of the matter superfields, and $C(R)$ is given by

\begin{equation}\label{C(R)}
\mbox{tr}\,(T^a T^b) = C(R)\,\delta^{ab}.
\end{equation}

\noindent In order to obtain such $\beta$-function it is most
convenient to use the higher derivative regularization
\cite{Slavnov,Bakeyev}. Then the renormalization of the operator
$W_a C^{ab} W_b$ is exhausted at the one-loop, while the
Gell-Mann-Low function coincides with the exact NSVZ
$\beta$-function. This was confirmed by the explicit calculations
in two-loop \cite{hep,tmf2}, three-loop \cite{ThreeLoop}, and
partially four-loop approximations \cite{Pimenov}. (We note that
the Gell-Mann-Low function does not depend on a particular choice
of the renormalization scheme due to its definition.)

Investigation of the $N=1$ supersymmetric electrodynamics was made
in Ref. \cite{SD} exactly to all orders of the perturbation
theory. According to that paper, it is possible to obtain the
exact $\beta$-function using Schwinger-Dyson equations and Ward
identities. However, there is an additional contribution besides
the ordinary NSVZ result. Explicit calculations show that it is
always equal to 0 in the lowest orders. This allows proposing
existence of a new identity for some Green functions. A proof of
this identity exactly to all orders is rather nontrivial. The
matter is that it is not reduced to the Ward identity, while its
proving by summation of diagrams \cite{Identity} is rather
complicated. We note that in Ref. \cite{Identity} this proof was
made only for a restricted class of diagrams in the massless case.
Nevertheless, it seems that the new identity is also valid for the
other diagrams, that follows, for example, from the explicit
four-loop calculations \cite{Pimenov}.

So, obtaining the exact $\beta$-function is a rather nontrivial
problem even in the electrodynamics. In the case of the Yang-Mills
theory the exact $\beta$-function agrees with the results of
explicit calculations in the two-loop approximation if the
dimensional reduction is used for regularization. However, there
are disagreements between the predictions, following from the
exact $\beta$-function, and explicit calculations, made in
$\overline{\mbox{MS}}$-scheme for the $\beta$-function, defined as
a derivative of the renormalized coupling constant, in the
three-loop approximation \cite{ThreeLoop1,ThreeLoop2,ThreeLoop3}.
It was shown \cite{ThreeLoop2} that these disagreements can be
eliminated by a special choice of renormalization scheme, the
possibility of such a choice being highly nontrivial
\cite{JackJones}.

In this paper we attempt to investigate contribution of the matter
superfields to the Gell-Mann-Low function in non-Abelian
supersymmetric gauge theories exactly to all orders of the
perturbation theory using the Schwinger-Dyson equations and
Slavnov-Taylor identities. The Gell-Mann-Low function is defined
by asymptotic of the two-point Green function in the limit of
large momentums. And so, the obtained results actually correspond
to the limit of massless matter fields. (Nevertheless, in order to
elucidate calculating the contribution of Pauli-Villars fields, we
will consider the case of an arbitrary mass throughout the paper.
Taking the limit $m\to 0$ will be made in the end for obtaining
the final result.)

This paper is organized as follows.

In Sec. \ref{Section_SUSY_QED} we recall basic information about
the $N=1$ supersymmetric Yang-Mills theory and its higher
derivatives regularization. The Schwinger-Dyson equations for the
considered theory are obtained in Sec. \ref{Section_SD}. In that
section we present the solution of the Slavnov-Taylor identities,
which determines vertex functions in the Schwinger-Dyson
equations. After substitution of these vertexes the contribution
to the two-point Green function of the gauge field is obtained in
the form of an integral. Calculation of this integral is made in
Sec. \ref{Section_Two_Point_Function}. A brief discussion of the
results is given in the conclusion.


\section{$N=1$ supersymmetric Yang-Mills theory, background field
method and higher derivative regularization}
\label{Section_SUSY_QED} \hspace{\parindent}

The $N=1$ supersymmetric Yang-Mills theory in the superspace is
described by the action

\begin{eqnarray}\label{SYM_Action}
&& S = \frac{1}{4 e^2} \mbox{Re}\,\mbox{tr}\int
d^4x\,d^2\theta\,W_a C^{ab} W_b + \frac{1}{4}\int d^4x\,
d^4\theta\, \Big(\phi^+ e^{2V}\phi
+\tilde\phi^+ e^{-2V^{t}}\tilde\phi\Big) +\nonumber\\
&& + \frac{1}{2}m\int d^4x\, d^2\theta\,\tilde\phi^t\,\phi +
\frac{1}{2}m\int d^4x\, d^2\bar\theta\,\tilde\phi^+\phi^*.
\end{eqnarray}

\noindent Here $\phi$ and $\tilde\phi$ are chiral matter
superfields, and $V$ is a real scalar superfield, which contains
the gauge field $A_\mu$ as a component. The superfield $W_a$ is a
supersymmetric analogue of the gauge field stress tensor. It is
defined by

\begin{equation}
W_a = \frac{1}{32} \bar D (1-\gamma_5) D\Big[e^{-2V}
(1+\gamma_5)D_a e^{2V}\Big],
\end{equation}

\noindent where

\begin{equation}
D = \frac{\partial}{\partial\bar\theta} -
i\gamma^\mu\theta\,\partial_\mu
\end{equation}

\noindent is a supersymmetric covariant derivative. In our
notation, the gauge superfield $V$ is expanded over the generators
of the gauge group $T^a$ as $V = e\, V^a T^a$, where $e$ is a
coupling constant.

Action (\ref{SYM_Action}) is invariant under gauge transformations

\begin{equation}
\phi \to e^{i\Lambda}\phi;\qquad \tilde\phi \to
e^{-i\Lambda^t}\tilde\phi;\qquad e^{2V} \to e^{i\Lambda^+} e^{2V}
e^{-i\Lambda}.
\end{equation}

\noindent Such a transformation law means that if the field $\phi$
is in the representation $R$ of the gauge group $G$, the field
$\tilde\phi$ is in the representation $\bar R$, conjugated to $R$.

For quantization of this model it is convenient to use the
background field method. The matter is that the background field
method allows calculating the effective action without manifest
breaking of the gauge invariance. In the supersymmetric case it
can formulated as follows \cite{West, Superspace}: Let us make a
substitution

\begin{equation}\label{Substitution}
e^{2V} \to e^{2V'} \equiv e^{\mbox{\boldmath${\scriptstyle
\Omega}$}^+} e^{2V} e^{\mbox{\boldmath${\scriptstyle \Omega}$}},
\end{equation}

\noindent in action (\ref{SYM_Action}), where
$\mbox{\boldmath${\Omega}$}$ is a background scalar superfield.
Expression for $V'$ is a complicated nonlinear function of $V$,
$\mbox{\boldmath$\Omega$}$, and $\mbox{\boldmath$\Omega^+$}$. We
do not interested in explicit form of this function:

\begin{equation}
V' = V'[V,\mbox{\boldmath$\Omega$}].
\end{equation}

\noindent (For brevity of notation we will not explicitly write
the dependence on $\mbox{\boldmath$\Omega$}^+$ here and below.)

The obtained theory will be invariant under background gauge
transformations

\begin{eqnarray}\label{Background_Transformations}
&& \phi \to e^{\Lambda}\phi;\qquad \tilde\phi \to
e^{-\Lambda^t}\tilde\phi;\qquad V \to e^{iK} V e^{-iK};
\nonumber\\
&& e^{\mbox{\boldmath${\scriptstyle \Omega}$}} \to e^{iK}
e^{\mbox{\boldmath${\scriptstyle \Omega}$}} e^{-\Lambda};\qquad
e^{\mbox{\boldmath${\scriptstyle \Omega}$}^+} \to e^{-\Lambda^+}
e^{\mbox{\boldmath${\scriptstyle \Omega}$}^+} e^{-iK},
\end{eqnarray}

\noindent where $K$ is a real superfield, and $\Lambda$ is a
chiral superfield.

Let us construct the chiral covariant derivatives

\begin{equation}
\mbox{\boldmath$D$} \equiv e^{-\mbox{\boldmath${\scriptstyle
\Omega}$}^+} \frac{1}{2} (1+\gamma_5)D
e^{\mbox{\boldmath${\scriptstyle \Omega}$}^+};\qquad
\bar{\mbox{\boldmath$D$}} \equiv e^{\mbox{\boldmath${\scriptstyle
\Omega}$}} \frac{1}{2} (1-\gamma_5) D
e^{-\mbox{\boldmath${\scriptstyle \Omega}$}}.
\end{equation}

\noindent Acting on some field $X$, which is transformed as $X \to
e^{iK} X$, these covariant derivatives are transformed in the same
way. It is also possible to define a covariant derivative with the
Lorentz index

\begin{equation}
\mbox{\boldmath$D$}_\mu \equiv - \frac{i}{4} (C\gamma^\mu)^{ab}
\Big\{\mbox{\boldmath$D$}_a,\bar{\mbox{\boldmath$D$}}_b\Big\},
\end{equation}

\noindent which will have the same property. It is easy to see
that after substitution (\ref{Substitution}) action
(\ref{SYM_Action}) will be

\begin{eqnarray}\label{Background_Action}
&& S = \frac{1}{4 e^2}\mbox{tr}\,\mbox{Re}\int d^4x\,d^2\theta\,
\mbox{\boldmath$W$}^a \mbox{\boldmath$W$}_a - \frac{1}{128
e^2}\mbox{tr}\,\mbox{Re}\int d^4x\,d^4\theta\,\Bigg[16
\Big(e^{-2V}\mbox{\boldmath$D$}^a e^{2V}\Big)
\mbox{\boldmath$W$}_a
+\nonumber\\
&& + \Big(e^{-2V}\mbox{\boldmath$D$}^a e^{2V}\Big)
\bar{\mbox{\boldmath$D$}}^2 \Big(e^{-2V}\mbox{\boldmath$D$}_a
e^{2V}\Big) \Bigg] + \frac{1}{4}\int d^4x\, d^4\theta\,
\Big(\phi^+ e^{\mbox{\boldmath${\scriptstyle \Omega}$}^+} e^{2V}
e^{\mbox{\boldmath${\scriptstyle \Omega}$}} \phi +\tilde\phi^+
e^{-\mbox{\boldmath${\scriptstyle \Omega}$}^*} e^{-2V^t}
\times\nonumber\\
&& \times e^{-\mbox{\boldmath${\scriptstyle \Omega}$}^t}
\tilde\phi\Big) + \frac{1}{2}m\int d^4x\,
d^2\theta\,\tilde\phi^t\,\phi + \frac{1}{2}m\int d^4x\,
d^2\bar\theta\,\tilde\phi^+\phi^*,
\end{eqnarray}

\noindent where

\begin{equation}
\mbox{\boldmath$W$}_a = \frac{1}{32}
e^{\mbox{\boldmath${\scriptstyle \Omega}$}} \bar D (1-\gamma_5) D
\Big(e^{-\mbox{\boldmath${\scriptstyle \Omega}$}}
e^{-\mbox{\boldmath${\scriptstyle \Omega}$}^+} (1+\gamma_5) D_a
e^{\mbox{\boldmath${\scriptstyle \Omega}$}^+}
e^{\mbox{\boldmath${\scriptstyle \Omega}$}} \Big)
e^{-\mbox{\boldmath${\scriptstyle \Omega}$}},
\end{equation}

\noindent and the notation

\begin{eqnarray}\label{Derivatives_Notations}
&& \mbox{\boldmath$D$}^2 \equiv \frac{1}{2} \bar
{\mbox{\boldmath$D$}} (1+\gamma_5)\mbox{\boldmath$D$};\qquad\ \bar
{\mbox{\boldmath$D$}}^2 \equiv \frac{1}{2} \bar
{\mbox{\boldmath$D$}} (1-\gamma_5) \mbox{\boldmath$D$};\nonumber\\
&& \mbox{\boldmath$D$}^a \equiv \Big[\frac{1}{2}\bar
{\mbox{\boldmath$D$}} (1+\gamma_5)\Big]^a;\qquad
\mbox{\boldmath$D$}_a \equiv \Big[\frac{1}{2}(1+\gamma_5)
\mbox{\boldmath$D$}\Big]_a;\nonumber\\
&& \bar {\mbox{\boldmath$D$}}^a \equiv \Big[\frac{1}{2}\bar
{\mbox{\boldmath$D$}} (1 - \gamma_5)\Big]^a;\qquad \bar
{\mbox{\boldmath$D$}}_a \equiv \Big[\frac{1}{2}(1-\gamma_5)
\mbox{\boldmath$D$}\Big]_a
\end{eqnarray}

\noindent is used. Action of the covariant derivatives on the
field $V$ in the adjoint representation is defined by the standard
way.

It is convenient to choose a regularization and gauge fixing so
that invariance (\ref{Background_Transformations}) will be
unbroken. For example, the higher covariant derivative
regularization can be introduced by the substitution of action
(\ref{Background_Action}) for

\begin{eqnarray}\label{Regularized_Action}
&& S_{\Lambda} = \frac{1}{4 e^2}\mbox{tr}\,\mbox{Re}\int
d^4x\,d^2\theta\, \mbox{\boldmath$W$}^a \mbox{\boldmath$W$}_a -
\frac{1}{128 e^2}\mbox{tr}\,\mbox{Re}\int
d^4x\,d^4\theta\,\Bigg[16 \Big(e^{-2V}\mbox{\boldmath$D$}^a
e^{2V}\Big) \mbox{\boldmath$W$}_a
+\nonumber\\
&& + \Big(e^{-2V}\mbox{\boldmath$D$}^a e^{2V}\Big)
\bar{\mbox{\boldmath$D$}}^2 \Big(e^{-2V}\mbox{\boldmath$D$}_a
\Big(1 + \frac{(\mbox{\boldmath$D$}_\mu^2)^n}{\Lambda^{2n}}\Big)
e^{2V}\Big) \Bigg]+ \frac{1}{4}\int d^4x\, d^4\theta\, \Big(\phi^+
e^{\mbox{\boldmath${\scriptstyle \Omega}$}^+} e^{2V}
e^{\mbox{\boldmath${\scriptstyle \Omega}$}} \phi +\nonumber\\
&& +\tilde\phi^+ e^{-\mbox{\boldmath${\scriptstyle \Omega}$}^*}
e^{-2V^t} e^{-\mbox{\boldmath${\scriptstyle \Omega}$}^t}
\tilde\phi\Big) + \frac{1}{2}m\int d^4x\,
d^2\theta\,\tilde\phi^t\,\phi + \frac{1}{2}m\int d^4x\,
d^2\bar\theta\,\tilde\phi^+\phi^*.
\end{eqnarray}

\noindent This action is evidently invariant under background
gauge transformations (\ref{Background_Transformations}).

We note that the described way of regularization is a bit
different from the method, proposed in Ref. \cite{West_Paper}. The
difference is in the form of the term, which contains higher
derivatives. Using of the method, proposed here, simplifies
calculations in a certain degree, while all particular features of
higher derivative regularization are the same in the both cases.
So, although explicit calculations by the method, proposed in Ref.
\cite{West_Paper} in two and more loops were not yet made, it is
possible to propose that their results will not essentially differ
from the ones, obtained below in this paper.

The gauge can be also fixed by the invariant way, for example, by
adding the terms

\begin{equation}\label{Gauge_Fixing}
S_{gf} = - \frac{1}{64 e^2}\int d^4x\,d^4\theta\, \Bigg(V
\mbox{\boldmath$D$}^2 \bar{\mbox{\boldmath$D$}}^2 \Big(1 +
\frac{(\mbox{\boldmath$D$}_\mu^2)^n}{\Lambda^{2n}}\Big) V + V \bar
{\mbox{\boldmath$D$}}^2 \mbox{\boldmath$D$}^2 \Big(1+
\frac{(\mbox{\boldmath$D$}_\mu^2)^n}{\Lambda^{2n}}\Big) V\Bigg).
\end{equation}

\noindent In this case terms quadratic in the superfield $V$ will
have the simplest form:

\begin{equation}
\frac{1}{4 e^2}\mbox{tr}\,\mbox{Re}\int d^4x\,d^4\theta\, V
\mbox{\boldmath$D$}_\mu^2 \Big(1+
\frac{(\mbox{\boldmath$D$}_\mu^2)^n}{\Lambda^{2n}}\Big) V.
\end{equation}

\noindent Certainly, it is also necessary to add the corresponding
action $S_{gh}$ for the Faddeev-Popov ghosts. We will not write
here explicit form of this action, because it is not essential for
calculating a contribution of the matter superfield loops.

Proposed way of the regularization and gauge fixing preserves both
invariance under the supersymmetry transformations and the
invariance under transformations
(\ref{Background_Transformations}). As a consequence, the
effective action, calculated by the background field method, will
be invariant under both supersymmetry and gauge transformations.

Let us construct the generating functional as follows:

\begin{eqnarray}\label{Generating_Functional}
&& Z[J,\mbox{\boldmath$\Omega$},j] = \int D\mu\,\exp\Big\{i
S_\Lambda^r  + i S_{gf} + i S_{gh} + i S_{S} + i S_{\phi_0}
+\nonumber\\
&& \qquad\qquad\qquad\qquad\qquad + i \int d^4x\,d^4\theta\,\Big(J
+ J[\mbox{\boldmath$\Omega$}] \Big)
\Big(V'[V,\mbox{\boldmath$\Omega$}] - {\bf V} \Big) \Big\},\qquad
\end{eqnarray}

\noindent where the superfield ${\bf V}$ is given by

\begin{equation}\label{Background Field}
e^{2{\bf V}} \equiv e^{\mbox{\boldmath${\scriptstyle \Omega}$}^+}
e^{\mbox{\boldmath${\scriptstyle \Omega}$}},
\end{equation}

\noindent and $J[\mbox{\boldmath$\Omega$}]$ is a so far undefined
functional. A reason of its introducing will be clear later. The
functional integration measure is written as

\begin{equation}
D\mu = DV\,D\phi\,D\tilde\phi.
\end{equation}

\noindent $S_\Lambda^r$ denotes renormalized and regularized
action

\begin{eqnarray}\label{Renormalized_Action}
&& S_{\Lambda}^r = \frac{1}{4 e_0^2}\mbox{tr}\,\mbox{Re}\int
d^4x\,d^2\theta\, \mbox{\boldmath$W$}^a \mbox{\boldmath$W$}_a -
\frac{1}{128 e_0^2}\mbox{tr}\,\mbox{Re}\int
d^4x\,d^4\theta\,\Bigg[16 \Big(e^{-2V}\mbox{\boldmath$D$}^a
e^{2V}\Big) \mbox{\boldmath$W$}_a
+\nonumber\\
&& + \Big(e^{-2V}\mbox{\boldmath$D$}^a e^{2V}\Big)
\bar{\mbox{\boldmath$D$}}^2 \Big(e^{-2V}\mbox{\boldmath$D$}_a
\Big(1 + \frac{(\mbox{\boldmath$D$}_\mu^2)^n}{\Lambda^{2n}}\Big)
e^{2V}\Big) \Bigg]+ \frac{1}{4} Z(e,\Lambda/\mu) \int d^4x\,
d^4\theta\, \Big(\phi^+ e^{\mbox{\boldmath${\scriptstyle
\Omega}$}^+} \times\nonumber\\
&& \times e^{2V} e^{\mbox{\boldmath${\scriptstyle \Omega}$}} \phi
+\tilde\phi^+ e^{-\mbox{\boldmath${\scriptstyle \Omega}$}^*}
e^{-2V^t} e^{-\mbox{\boldmath${\scriptstyle \Omega}$}^t}
\tilde\phi\Big) + \frac{1}{2}m\int d^4x\,
d^2\theta\,\tilde\phi^t\,\phi + \frac{1}{2}m\int d^4x\,
d^2\bar\theta\,\tilde\phi^+\phi^*,
\end{eqnarray}

\noindent where $e_0$ is a bare coupling constant, and $Z$ is a
renormalization constant for the wave function of the matter
superfields. $S_{gf}$ denotes gauge fixing terms
(\ref{Gauge_Fixing}) (the coupling constant $e$ in them is also
replaced by $e_0$), and $S_{gh}$ is a corresponding action for the
Faddeev-Popov ghosts (and also for the Nielsen-Kallosh ghosts).
$S_S$ denotes the terms with sources for chiral superfields, which
in the extended form are written as

\begin{equation}\label{Sources}
S_S = \int d^4x\,d^2\theta\, \Big(j^t\,\phi + \tilde
j^t\,\tilde\phi \Big) + \int d^4x\,d^2\bar\theta\, \Big(j^+\phi^*
+ \tilde j^+ \tilde\phi^*\Big).
\end{equation}

\noindent Moreover, in generating functional
(\ref{Generating_Functional}) we introduce the supplementary
sources

\begin{eqnarray}\label{New_Sources}
&& S_{\phi_0} = \frac{1}{4}\int d^4x\,d^4\theta\,\Big(\phi_0^+
e^{\mbox{\boldmath${\scriptstyle \Omega}$}^+} e^{2V}
e^{\mbox{\boldmath${\scriptstyle \Omega}$}} \phi + \phi^+
e^{\mbox{\boldmath${\scriptstyle \Omega}$}^+} e^{2V}
e^{\mbox{\boldmath${\scriptstyle \Omega}$}}
\phi_0 +\nonumber\\
&& \qquad\qquad\qquad\qquad + \tilde\phi_0^+
e^{-\mbox{\boldmath${\scriptstyle \Omega}$}^*} e^{-2V^t}
e^{-\mbox{\boldmath${\scriptstyle \Omega}$}^t} \tilde\phi +
\tilde\phi^+ e^{-\mbox{\boldmath${\scriptstyle \Omega}$}^*}
e^{-2V^t} e^{-\mbox{\boldmath${\scriptstyle \Omega}$}^t}
\tilde\phi_0 \Big),\qquad
\end{eqnarray}

\noindent where $\phi_0$, $\phi_0^+$, $\tilde\phi_0$ and
$\tilde\phi_0^+$ are arbitrary scalar superfields. In principle,
it is not necessary to introduce the term $S_{\phi_0}$ in the
generating functional, but the presence of the parameters $\phi_0$
is highly desirable for investigating the Schwinger-Dyson
equations.

In order to understand how generating functional
(\ref{Generating_Functional}) is related with the ordinary
effective action, we perform the substitution $V \to V'$. Then we
obtain

\begin{equation}
Z[J,\mbox{\boldmath$\Omega$},j] = \exp\Big\{ -i \int
d^4x\,d^4\theta\,\Big(J + J[\mbox{\boldmath$\Omega$}]\Big) {\bf V}
\Big\} Z_0\Big[J + J[\mbox{\boldmath$\Omega$}],
\mbox{\boldmath$\Omega$},j\Big],
\end{equation}

\noindent where

\begin{equation}
Z_0[J,\mbox{\boldmath$\Omega$},j] = \int D\mu\,\exp\Big\{i
S_\Lambda^r + i S_{gf} + i S_{gh} + i S_S + i S_{\phi_0} + i \int
d^4x\,d^4\theta\,J V \Big\}.
\end{equation}

\noindent If the dependence of $S_\Lambda^r$, $S_{gf}$, $S_{gh}$,
and $S_{\phi_0}$ on the arguments $V$, $\mbox{\boldmath$\Omega$}$,
and $\mbox{\boldmath$\Omega^+$}$ were factorized into the
dependence on the variable $V'$, $Z_0$ would not depend on
$\mbox{\boldmath$\Omega$}$ and $\mbox{\boldmath$\Omega^+$}$ and
would coincide with the ordinary generating functional. This
really takes place for action (\ref{SYM_Action}) and $S_{\phi_0}$.
However, in the term with the higher derivatives and in the gauge
fixing terms such factorization does not occur. Therefore, $Z_0$
actually differs from the ordinary generating functional.

Using the functional $Z[J,\mbox{\boldmath$\Omega$},j]$ it is
possible to construct the generating functional for the connected
Green functions

\begin{eqnarray}
&& W[J,\mbox{\boldmath$\Omega$},j] = -i\ln
Z[J,\mbox{\boldmath$\Omega$},j]
=\nonumber\\
&&\qquad\qquad = - \int d^4x\,d^4\theta \Big(J +
J[\mbox{\boldmath$\Omega$}]\Big) {\bf V} + W_0\Big[J +
J[\mbox{\boldmath$\Omega$}],\mbox{\boldmath$\Omega$},j\Big].\qquad
\end{eqnarray}

\noindent Also it is possible to construct the corresponding
effective action

\begin{eqnarray}\label{Effective_Action_Definition}
&& \Gamma[V,\mbox{\boldmath$\Omega$}, \phi] = -\int
d^4x\,d^4\theta\,\Big(J {\bf V} + J[\mbox{\boldmath$\Omega$}] {\bf
V}\Big) + W_0\Big[J +
J[\mbox{\boldmath$\Omega$}],\mbox{\boldmath$\Omega$},j\Big] -
\nonumber\\
&& - \int d^4x\,d^4\theta\,J V - \int d^4x\,d^2\theta\,
\Big(j^t\,\phi + \tilde j^t\,\tilde\phi \Big) - \int
d^4x\,d^2\bar\theta\, \Big(j^+ \phi^* + \tilde j^+ \tilde\phi^*
\Big),
\end{eqnarray}

\noindent where the sources should be expressed in terms of fields
using the equation

\begin{eqnarray}
&& V = \frac{\delta}{\delta J} W[J,\mbox{\boldmath$\Omega$},j] = -
{\bf V} + \frac{\delta}{\delta J} W_0\Big[J +
J[\mbox{\boldmath$\Omega$}],\mbox{\boldmath$\Omega$},j\Big];
\nonumber\\
&& \phi = \frac{\delta}{\delta j} W[J,\mbox{\boldmath$\Omega$},j]
= \frac{\delta}{\delta j} W_0\Big[J +
J[\mbox{\boldmath$\Omega$}],\mbox{\boldmath$\Omega$},j\Big]\quad
\mbox{e.t.c.}
\end{eqnarray}

\noindent Substituting these expression into Eq.
(\ref{Effective_Action_Definition}), we write the effective action
as

\begin{eqnarray}
&& \Gamma[V,\mbox{\boldmath$\Omega$},\phi] = W_0\Big[J +
J[\mbox{\boldmath$\Omega$}],\mbox{\boldmath$\Omega$},j\Big] - \int
d^4x\,d^4\theta\,\Big(J[\mbox{\boldmath$\Omega$}] {\bf V} + J
\frac{\delta}{\delta J} W_0\Big[j,J +
J[\mbox{\boldmath$\Omega$}],\mbox{\boldmath$\Omega$},j\Big]\Big)
-\nonumber\\
&& - \int d^4x\,d^2\theta\,\phi\frac{\delta}{\delta j} W_0\Big[J +
J[\mbox{\boldmath$\Omega$}],\mbox{\boldmath$\Omega$},j\Big]
-\Big(\mbox{similar terms with $\phi^+$, $\tilde\phi$ and
$\tilde\phi^+$}\Big).
\end{eqnarray}

\noindent Let us now set $V = 0$, so that

\begin{equation}\label{Phi_To_Zero}
{\bf V} = \frac{\delta}{\delta J} W_0\Big[J +
J[\mbox{\boldmath$\Omega$}],\mbox{\boldmath$\Omega$},j\Big].
\end{equation}

Moreover, we take into account that the invariance under
background gauge transformations
(\ref{Background_Transformations}) essentially restricts the form
of the effective action. If the quantum field $V$ in the effective
action is set to 0, the superfield $K$ will affect only the gauge
transformation law of the fields $\mbox{\boldmath$\Omega$}$ and
$\mbox{\boldmath$\Omega$}^+$, and the only invariant combination
is expression (\ref{Background Field}). (It is invariant in a
sense, that the corresponding transformation law does not contain
the superfield $K$.) This means that in the final expression for
the effective action we can set

\begin{equation}\label{K_Fixing}
\mbox{\boldmath$\Omega$} = \mbox{\boldmath$\Omega$}^+ = {\bf V}.
\end{equation}

\noindent In this case the effective action is

\begin{eqnarray}\label{Background_Gamma}
&& \Gamma[0,{\bf V},\phi] = W_0\Big[J + J[{\bf V}],{\bf V},j\Big]
- \int d^4x\,d^4\theta\,\Big(J + J[{\bf V}]\Big)
\frac{\delta}{\delta J} W_0\Big[J + J[{\bf V}], {\bf V}, j\Big]
-\nonumber\\
&& - \int d^4x\,d^2\theta\,\phi\frac{\delta}{\delta j} W_0\Big[J +
J[{\bf V}],{\bf V},j\Big] -\Big(\mbox{similar terms with $\phi^+$,
$\tilde\phi$, and $\tilde\phi^+$}\Big).
\end{eqnarray}

\noindent Note, that this expression does not depend on form of
the functional $J[\mbox{\boldmath$\Omega$}]$. In particular it can
be chosen to cancel terms linear in the field $V$ in Eq.
(\ref{Generating_Functional}). Such a choice will be very
convenient below.

If the gauge fixing terms and the terms with higher derivatives
depended only on $V'$, expression (\ref{Background_Gamma}) would
coincide with the ordinary effective action. However, as we
already mentioned above, the dependence on $V$,
$\mbox{\boldmath$\Omega$}$, and $\mbox{\boldmath$\Omega$}^+$ is
not factorized into the dependence on $V'$ in the proposed method
of renormalization and gauge fixing. According to Ref.
\cite{Kluberg1,Kluberg2} the invariant charge (and, therefore, the
Gell-Mann-Low function) is gauge independent, and the dependence
of the effective action on gauge can be eliminated by
renormalization of the wave functions of the gauge field, ghosts,
and matter fields. Therefore, for calculating the Gell-Mann-Low
function we may use the background gauge described above. We note
that if this gauge is used, the renormalization constant of the
gauge field $A_\mu$ is 1 due to the invariance of the action under
transformations (\ref{Background_Transformations}).

Nevertheless, generating functional (\ref{Generating_Functional})
is not yet completely constructed. The matter is that adding the
term with higher derivatives does not remove divergences from
one-loop diagrams. To regularize them, it is necessary to insert
the Pauli-Villars determinants in the generating functional
\cite{Slavnov_Book}. Because in this paper we are interested only
in contributions of matter superfields, we construct these
determinants only for them. (Also it is necessary to introduce the
Pauli-Villars fields for the gauge field and ghosts.) Here we will
at once use condition (\ref{K_Fixing}). So, we insert in the
generating functional the factor

\begin{equation}\label{PV_Insersion}
\prod\limits_i \Big(\det PV(V,{\bf V},M_i)\Big)^{c_i},
\end{equation}

\noindent in which the Pauli-Villars determinants are defined by

\begin{equation}\label{PV_Determinants}
\Big(\det PV({\bf V},M)\Big)^{-1} = \int D\Phi\,D\tilde \Phi\,
\exp\Big(i S_{PV}\Big).
\end{equation}

\noindent If condition (\ref{K_Fixing}) is used, the action for
the Pauli-Villars fields will be

\begin{eqnarray}
&& S_{PV}\equiv Z(e,\Lambda/\mu) \frac{1}{4} \int
d^4x\,d^4\theta\, \Big(\Phi^+ e^{\bf V} e^{2V} e^{\bf V} \Phi +
\tilde\Phi^+ e^{-{\bf V}^t} e^{-2V^t} e^{-{\bf V}^t}\tilde\Phi
\Big)
+\qquad\nonumber\\
&& + \frac{1}{2}\int d^4x\,d^2\theta\, M \tilde\Phi^t \Phi +
\frac{1}{2}\int d^4x\,d^2\bar\theta\, M \tilde\Phi^+ \Phi^*.\qquad
\end{eqnarray}

\noindent The coefficients $c_i$ in Eq. (\ref{PV_Insersion})
satisfy conditions

\begin{equation}
\sum\limits_i c_i = 1;\qquad \sum\limits_i c_i M_i^2 = 0.
\end{equation}

\noindent Below, we assume that $M_i = a_i\Lambda$, where $a_i$
are some constants. Inserting the Pauli-Villars determinants
allows cancelling the remaining divergences in all one-loop
diagrams, including diagrams containing counterterm insertions.

Making the transformation

\begin{equation}
\phi \to \phi/\sqrt{Z};\quad \tilde\phi \to
\tilde\phi/\sqrt{Z};\quad \Phi \to \Phi/\sqrt{Z};\quad \tilde\Phi
\to \tilde\Phi/\sqrt{Z}
\end{equation}

\noindent in the generating functional, we obtain that the
dependence on the renormalization constant $Z$ can be easily found
from the generating functional, which does not contain the
constant $Z$. If $\phi_0$ and $j$ are 0, this is made by the
substitution

\begin{equation}\label{M_Substitution}
m \to m/Z;\qquad M_i \to M_i/Z.
\end{equation}

\noindent Therefore, we will make calculations with $Z=1$, and
dependence on the renormalization constant will be restored in the
final result.

\section{Schwinger-Dyson equations and Slavnov-Taylor identities.}
\hspace{\parindent} \label{Section_SD}

In order to construct the Schwinger-Dyson equations for the
considered theory it is necessary to split the action into three
parts: the action for the background field, the kinetic term for
quantum fields, which does not contain the background field, and
interaction, in which the other terms are included:

\begin{equation}
S = S({\bf V}) + S_2(V,\phi) + S_I(V,{\bf V},\phi).
\end{equation}

\noindent (Earlier we saw that the terms of the first order in the
superfield $V$, which were obtained from the expansion of the
classical action, can be omitted.) So, generating functional
(\ref{Generating_Functional}) can be written as

\begin{eqnarray}
&& Z[J,j,{\bf V}] = \int d\mu\,\exp\Big(i S[{\bf V}] + i
S_2[V,\phi] + i S_I[V,{\bf V},\phi] + i S_S + i \int d^8x\,J V
\Big)
=\\
&& = \exp\Big(i S[{\bf V}]+iS_I\Big[\frac{1}{i}
\frac{\delta}{\delta J},\frac{1}{i} \frac{\delta}{\delta {\bf J}},
\frac{1}{i}\frac{\delta}{\delta j} \Big]\Big) \times\nonumber\\
&& \qquad\qquad\qquad\qquad \times \int d\mu\,\exp\Big(i
S_2[V,\phi] + i S_S + i \int d^8x\,J V + i \int d^8x\,{\bf J} {\bf
V}\Big)\Bigg|_{{\bf J}=0},\qquad\nonumber
\end{eqnarray}

\noindent where ${\displaystyle \int d^8x \equiv \int
d^4x\,d^4\theta_x}$. Let us differentiate this expression with
respect to the background field

\begin{eqnarray}
&& \frac{\delta}{\delta {\bf V}_x} Z[J,j,{\bf V}] = i \frac{\delta
S[{\bf V}]}{\delta {\bf V}}\,Z +
\exp\Big(iS_I\Big[\frac{1}{i}\frac{\delta}{\delta
J},\frac{1}{i}\frac{\delta}{\delta {\bf J}},
\frac{1}{i}\frac{\delta}{\delta j} \Big]\Big) i {\bf J_x}
\times\nonumber\\
&& \qquad\qquad\qquad\times \int d\mu\,\exp\Big(i S_2[V,\phi] + i
S_S + i \int d^8x\, J V + i \int d^8x\,{\bf J} {\bf
V}\Big)\Bigg|_{{\bf J}=0}.\qquad
\end{eqnarray}

\noindent Moving the current ${\bf J}_x$ to the left and dividing
the result to $Z$, we obtain

\begin{equation}
\frac{\delta}{\delta {\bf V}_x} W[J,j,{\bf V}] = \frac{\delta
S[{\bf V}]}{\delta {\bf V}_x} + \frac{\delta}{\delta {\bf V}_x}
S_I\Big[{\bf V}, V + \frac{1}{i}\frac{\delta}{\delta J},
\phi+\frac{1}{i}\frac{\delta}{\delta j}\Big]
\end{equation}

\noindent Because the background field ${\bf V}$ is a parameter of
the effective action, these equality can be equivalently written
as

\begin{equation}
\frac{\delta\Gamma}{\delta {\bf V}_x} = \frac{\delta S[{\bf
V}]}{\delta {\bf V}_x} + \Big\langle \frac{\delta}{\delta {\bf
V}_x} S_I\Big[{\bf V}, V, \phi\Big]\Big\rangle
\end{equation}

\noindent where the angular brackets denote taking an expectation
value by the ordinary functional integration. Certainly, it is
necessary to set the field $V$ (an argument of the effective
action) to 0 in the final result in accordance with Eq.
(\ref{Phi_To_Zero}) and the corresponding discussion.

Let us find a contribution to this expression given by the matter
superfields. The corresponding interaction terms are

\begin{equation}
S_I = \frac{1}{4} \int d^8x\,\phi^+ \Big( e^{\bf V} e^{2V} e^{\bf
V} -1\Big) \phi + \mbox{similar terms with }\tilde\phi.
\end{equation}

\noindent Differentiating $S_I$ with respect to the background
field, we obtain that the corresponding contribution to the
effective action is written in the form

\begin{eqnarray}\label{SD_SYM}
&& \frac{\delta\Gamma}{\delta {\bf V}_x^a} = \mbox{terms without
matter} + \sum\limits_i c_i \frac{\delta}{\delta {\bf
V}_x^a}\Big\langle\ln \det PV\Big(V,{\bf V},M_i\Big)\Big\rangle
+ \qquad\nonumber\\
&& + \frac{1}{4} \frac{\delta}{\delta {\bf V}_x^a} \int
d^8x\,\Big\langle \phi_x^+  e^{{\bf V}_x} e^{2V_x} e^{{\bf V}_x}
\phi_x + \mbox{similar terms with }\tilde\phi \Big\rangle.\quad
\end{eqnarray}

\noindent We are interested in the two-point Green function of the
gauge field, corresponding to the expansion of the effective
action in powers of the background field up to the second order
terms. To calculate such expansion, it is convenient to use the
following equality

\begin{eqnarray}\label{Expansion}
&& \int d^8x\,d^8y\,{\bf V}_x^a {\bf V}_y^b\frac{\delta^2}{\delta
{\bf V}_x^a \delta {\bf V}_y^b} \Big(e^{\bf V} e^{2V} e^{\bf
V}\Big)\Bigg|_{{\bf V}=0} = {\bf V} \Big({\bf V} e^{2V} +
e^{2V}{\bf V}\Big) + \Big({\bf V} e^{2V} +
e^{2V}{\bf V}\Big){\bf V} =\nonumber\\
&& = e \int d^8x\,d^8y\,{\bf V}_y^b {\bf V}_x^a \Bigg[ T^a
\frac{\delta}{\delta {\bf V}_y^b}\Big(e^{\bf V} e^{2V} e^{\bf
V}\Big)\Bigg|_{{\bf V}=0} + \frac{\delta}{\delta {\bf
V}_y^b}\Big(e^{\bf V} e^{2V} e^{\bf V}\Big)\Bigg|_{{\bf V}=0} T^a
\Bigg],
\end{eqnarray}

\noindent where $e$ in the beginning of the second string is a
coupling constant. Using the form of the action for supplementary
sources (\ref{New_Sources}), we easily obtain

\begin{equation}
\frac{1}{4} \Big\langle \phi_x^+ T^a e^{{\bf V}_x} e^{2V_x}
e^{{\bf V}_x} \phi_x + \phi_x^+ e^{{\bf V}_x} e^{2V_x} e^{{\bf
V}_x} T^a \phi_x \Big\rangle = \frac{1}{i}\mbox{tr}\Bigg[ T^a
\Bigg(\frac{\delta^2 \Gamma}{\delta j_x^+ \delta\phi_{0x}^+} +
\frac{\delta^2\Gamma}{\delta j_x \delta\phi_{0x}} \Bigg) \Bigg].
\end{equation}

\noindent (In the last equation the matrix notation is used for
the brevity.) The derivatives with respect to the sources must be
expressed in terms of fields as follows:

\begin{eqnarray}\label{Derivative}
&& \frac{\delta}{\delta j^+_x} = \int
d^8z\,\Bigg[\Bigg(\frac{\delta^2\Gamma}{\delta \phi_z \delta
\phi^+_x}\Bigg)^{-1} \,\frac{D^2}{8\partial^2}\frac{\delta}{\delta
\phi_z} + \Bigg(\frac{\delta^2\Gamma}{\delta \phi_z^+ \delta
\phi^+_x}\Bigg)^{-1}\,\frac{\bar
D^2}{8\partial^2}\frac{\delta}{\delta
\phi_z^+} +\nonumber\\
&& \qquad + \Bigg(\frac{\delta^2\Gamma}{\delta \tilde \phi_z
\delta \phi^+_x}\Bigg)^{-1}
\,\frac{D^2}{8\partial^2}\frac{\delta}{\delta \tilde \phi_z} +
\Bigg(\frac{\delta^2\Gamma}{\delta \tilde \phi_z^+ \delta
\phi^+_x}\Bigg)^{-1} \,\frac{\bar
D^2}{8\partial^2}\frac{\delta}{\delta \tilde\phi_z^+}+
\Bigg(\frac{\delta^2\Gamma}{\delta V_z \delta
\phi^+_x}\Bigg)^{-1}\frac{\delta}{\delta V_z}\Bigg].\qquad
\end{eqnarray}

Therefore, using Eqs. (\ref{SD_SYM}), (\ref{Expansion}), and
taking into account similar terms with the fields $\tilde\phi$,
the corresponding contribution to the two-point Green function of
the matter superfield can be written as

\begin{equation}\label{SD1}
\frac{\delta^2\Gamma}{\delta {\bf V}^b_y \delta {\bf V}^a_x} =
\ldots + e\,\frac{\delta}{\delta {\bf V}^b_y} \mbox{tr} \Bigg[T^a
\frac{1}{i}\frac{\delta^2 \Gamma}{\delta j_x^+\,
\delta\phi_{0x}^+} - (T^a)^t \frac{1}{i}
\frac{\delta^2\Gamma}{\delta\tilde j_x^+\,\delta\tilde\phi_{0x}^+}
+ \mbox{h.c.}\Bigg],
\end{equation}

\noindent where dots denote contributions of the gauge fields,
ghosts, and also all possible Pauli-Villars fields. (In this
expression we omit symmetrization with respect to the indexes $a$
and $b$, because, as we will see below, it will be symmetric
automatically.) We note that the calculation of the Pauli-Villars
fields contributions are made completely similar to the
calculation of ordinary fields contributions \cite{SD}, and
details of this calculation are not presented here. The result
will be given below.

In order to calculate expressions in Eq. (\ref{SD1}), it is
necessary to use Eq. (\ref{Derivative}) and manifestly perform the
differentiation. Then the result can be graphically presented as a
sum of two effective diagrams

\begin{equation}\label{SD_Equation}
\begin{picture}(0,1.8)
\put(-4.9,0.6){$\Delta\Gamma^{(2)}_V =$} \put(0.8,0.6){+}
\hspace*{-3.5cm}
\includegraphics[scale=0.88]{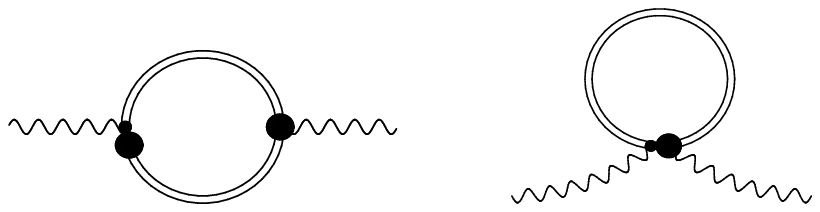}
\end{picture}
\end{equation}

\noindent The double lines correspond to the effective
propagators, which are written as

\begin{equation}\label{Inverse_Functions}
\Bigg(\frac{\delta^2\Gamma}{\delta\phi_x^+\delta\phi_y}\Bigg)^{-1}
= -  \frac{G D_x^2 \bar D_x^2}{4(\partial^2 G^2 + m^2 J^2)}
\delta^8_{xy};\qquad
\Bigg(\frac{\delta^2\Gamma}{\delta\phi_x\delta\tilde\phi_y}\Bigg)^{-1}
=  - \frac{m J \bar D_x^2}{\partial^2 G^2 + m^2 J^2}
\delta^8_{xy},
\end{equation}

\noindent depending on the chirality of the ends. The functions
$G$ and $J$ are determined by the two-point Green functions of the
matter superfield as

\begin{equation}\label{Explicit_Green_Functions}
\frac{\delta^2\Gamma}{\delta\phi_x^+\delta\phi_y} = \frac{D_x^2
\bar D_x^2}{16} G(\partial^2) \delta^8_{xy};\qquad
\frac{\delta^2\Gamma}{\delta\phi_x\delta\tilde\phi_y} = -
\frac{\bar D_x^2}{4} m J(\partial^2) \delta^8_{xy},
\end{equation}

\noindent where $\delta^8_{xy}\equiv
\delta^4(x-y)\delta^4(\theta_x-\theta_y)$, and the subscripts
denote points where the expressions are evaluated.

The vertex functions can be obtained from the Slavnov-Taylor
identities similar to the case of the electrodynamics. Certainly,
it is necessary to take into account that we consider the vertex
functions, which have an external line of the {\bf background}
field. As we already mentioned above, the effective action is
invariant under the background gauge transformations. It is easy
to see that this invariance can be expressed by the equality

\begin{eqnarray}\label{Invariance}
&& 0 = \int d^8y\,\Bigg(\frac{\delta\Gamma}{\delta {\bf
V}_y^a}\delta {\bf V}_y^a + \frac{\delta\Gamma}{\delta\phi_y}
\frac{D^2}{8\partial^2} \delta\phi_y + \delta\phi_y^+ \frac{\bar
D^2}{8\partial^2} \frac{\delta\Gamma}{\delta\phi_y^+} +
\frac{\delta\Gamma}{\delta\phi_{0y}} \delta\phi_{0y} +
\delta\phi_{0y}^+
\frac{\delta\Gamma}{\delta\phi_{0y}^+} +\qquad\nonumber\\
&& + \mbox{similar terms with $\tilde\phi$ and
$\tilde\phi_0$}\Bigg),
\end{eqnarray}

\noindent where

\begin{eqnarray}
&& \delta \phi_x = \Lambda_x \phi_x;\qquad \delta \phi_x^+ =
\phi_x^+ \Lambda_x^+;\qquad \delta \phi_{0x} = \Lambda_x
\phi_{0x};\qquad \delta \phi_{0x}^+ = \phi_{0x}^+ \Lambda_x^+;\nonumber\\
&& \delta {\bf V}_x = -\frac{1}{2}\Big(\Lambda_x +
\Lambda_x^+\Big) + O({\bf V}).
\end{eqnarray}

\noindent Here $\Lambda$ is an arbitrary chiral superfield, and
all other terms in $\delta {\bf V}$ are proportional at least to
the first degree of the background field. (For the fields
$\tilde\phi$ e.t.c. the transformation laws can be also easily
written.) Let us differentiate Eq. (\ref{Invariance}) with respect
to $\Lambda_y^a$, $\psi^+_z$, $\phi_x$ or with respect to
$\Lambda_y^a$, $\tilde\psi_z$, $\phi_x$. As a result we obtain the
Slavnov-Taylor identities

\begin{eqnarray}\label{ST_Identity}
&& 0 = \frac{1}{2e}(\bar D_y^2 + D_y^2)
\frac{\delta^3\Gamma}{\delta {\bf V}_y^a
\delta\phi^+_{0z}\delta\phi_x} - \frac{\delta^2\Gamma}{
\delta\phi_{0z}^+\delta\phi_y} T^a \bar D_y^2 \delta^8_{xy} -
D_y^2 \Big(\delta^8_{yz} T^a \frac{\delta^2\Gamma}{
\delta\phi_{0y}^+\delta\phi_x}\Big);\nonumber\\
&& 0 = \frac{1}{2e}(\bar D_y^2 + \bar D_y^2)
\frac{\delta^3\Gamma}{\delta {\bf V}^a_y \delta\tilde \phi_{0z}
\delta\phi_{x}} - \frac{\delta^2\Gamma}{\delta\tilde\phi_{0z}
\delta\phi_y} T^a \bar D_y^2 \delta^8_{xy} + \bar D_y^2 \Big(
\delta^8_{yz} T^a
\frac{\delta^2\Gamma}{\delta\tilde\phi_{0y}\delta\phi_x}\Big).
\end{eqnarray}

\noindent where

\begin{equation}\label{Useful_Identities}
\frac{\delta^2\Gamma}{\delta\phi_y \delta \phi_{0z}^+} = -
\frac{1}{8} G(\partial^2) \bar D_y^2\delta^8_{yz};\qquad
\frac{\delta^2\Gamma}{\delta\tilde\phi_y^* \delta \phi_{0z}^+} =
\frac{m}{32\partial^2} \Big(J(\partial^2)-1\Big) D_y^2 \bar D_y^2
\delta^8_{yz},
\end{equation}

\noindent They differ from the similar identities for the
electrodynamics only in the presence of gauge group generators.
Therefore, the solution will be also similar

\begin{eqnarray}\label{Vertex3}
&& \frac{\delta^3\Gamma}{\delta {\bf
V}^a_y\delta\phi^+_{0z}\delta\phi_x}\Bigg|_{p=0} = e \Bigg[-2
\partial^2\Pi_{1/2}{}_y\Big(\bar D_y^2\delta^8_{xy}
\delta^8_{yz}\Big) F(q^2) + \frac{1}{8} D^b C_{bc} \bar
D_y^2\Big(\bar D_y^2\delta^8_{xy} D_y^c \delta^8_{yz} \Big) f(q^2)
-\vphantom{\frac{1}{2}}\nonumber\\
&& -\frac{1}{16} q^\mu G'(q^2) \bar D\gamma^\mu\gamma_5 D_y
\Big(\bar D_y^2\delta^8_{xy} \delta^8_{yz}\Big) -\frac{1}{4} \bar
D_y^2\delta^8_{xy} \delta^8_{yz}\, G(q^2)\Bigg] T^a;\\
\label{Vertex4} && \frac{\delta^3\Gamma}{\delta {\bf V}^a_y
\delta\tilde\phi_{0z} \delta\phi_x}\Bigg|_{p=0} = e
\Bigg[\,\frac{m}{32} D^b C_{bc} D_y^2 \Big(D_y^2 \bar
D_y^2\delta^8_{xy} D_y^c \delta^8_{yz}\Big) h(q^2) +\frac{m}{16}
J'(q^2) \Bigg( \bar D_y^2\delta^8_{xy} D_y^2 \delta^8_{yz}
-\nonumber\\
&& - D_y^2 \bar D_y^2 \delta^8_{xy} \delta^8_{yz} \Bigg)
+\frac{m}{16}\Bigg(\frac{J'(q^2)}{q^2} - \frac{J(q^2)-1}{q^4}
\Bigg) \Bigg(D_y^2 \bar D_y^2\delta^8_{xy} \frac{\bar D_y^2
D_y^2}{16} \delta^8_{yz} + D_y^2 \bar D_y^2 \delta^8_{xy} q^2
\delta^8_{yz} \Bigg)\Bigg] T^a.\nonumber\\
\end{eqnarray}

\noindent The primes denote derivatives with respect to $q^2$,

\begin{equation}
\Pi_{1/2} = - \frac{1}{8 \partial^2} D^a \bar D^2 D_a = -
\frac{1}{8 \partial^2} \bar D^a D^2 \bar D_a
\end{equation}

\noindent is a supersymmetric transverse projection operator,
notation for the derivatives are similar to
(\ref{Derivatives_Notations}), and the functions $F$, $f$ and $h$
can not be determined from the Slavnov-Taylor identities.

Functions (\ref{Vertex3}) and (\ref{Vertex4}) allow finding
ordinary Green functions by the identities

\begin{equation}\label{Psi_Phi_Identites}
- \frac{D_z^2}{2}\frac{\delta^3\Gamma}{\delta {\bf
V}^a_x\delta\phi_y\delta\phi^+_{0z}} =
\frac{\delta^3\Gamma}{\delta {\bf
V}^a_x\delta\phi_y\delta\phi^+_z};\qquad - \frac{\bar D_z^2}{2}
\frac{\delta^3\Gamma}{\delta {\bf V}^a_x \delta\phi_y
\delta\tilde\phi_{0z}} = \frac{\delta^3\Gamma}{\delta {\bf V}^a_x
\delta\phi_y \delta\tilde\phi_z}
\end{equation}

\noindent We note that the supplementary sources in Eq.
(\ref{New_Sources}) were specially introduced in order that such
identities take place. A proof of these relations can be made
completely similar to the case of the electrodynamics, considered
in \cite{SD} in details. Using Eqs. (\ref{Psi_Phi_Identites}) we
find

\begin{eqnarray}\label{Vertex1}
&& \frac{\delta^3\Gamma}{\delta {\bf V}^a_y \delta\phi^+_z
\delta\phi_x} \Bigg|_{p=0} = e
\Bigg[\partial^2\Pi_{1/2}{}_y\Big(\bar D_y^2\delta^8_{xy} D_y^2
\delta^8_{yz}\Big) F(q^2) +\nonumber\\
&& \qquad\qquad +\frac{1}{32} q^\mu G'(q^2) \bar
D\gamma^\mu\gamma_5 D_y \Big(\bar D_y^2\delta^8_{xy} D_y^2
\delta^8_{yz}\Big) + \frac{1}{8} \bar D_y^2\delta^8_{xy} D_y^2
\delta^8_{yz}\, G(q^2)\Bigg] T^a;\\
\label{Vertex2} && \frac{\delta^3\Gamma}{\delta {\bf V}^a_y
\delta\tilde\phi_{z} \delta\phi_x }\Bigg|_{p=0} = -\frac{e m}{32}
J'(q^2) \Bigg[ \bar D_y^2\delta^8_{xy} D_y^2 \bar D_y^2
\delta^8_{yz} - D_y^2 \bar D_y^2 \delta^8_{xy} \bar
D_y^2\delta^8_{yz} \Bigg] T^a.\qquad
\end{eqnarray}

Obtained expressions for the vertex functions allow calculating
all expressions, entering Eq. (\ref{SD1}) in the limit of zero
external momentum. For this purpose it is convenient to make the
following observation: The first diagram in Eq.
(\ref{SD_Equation}) can be equivalently presented in the form

\begin{eqnarray}\label{Diagram_Identity}
\hspace*{-2cm}
\begin{picture}(0,1.8)
\includegraphics[scale=0.41]{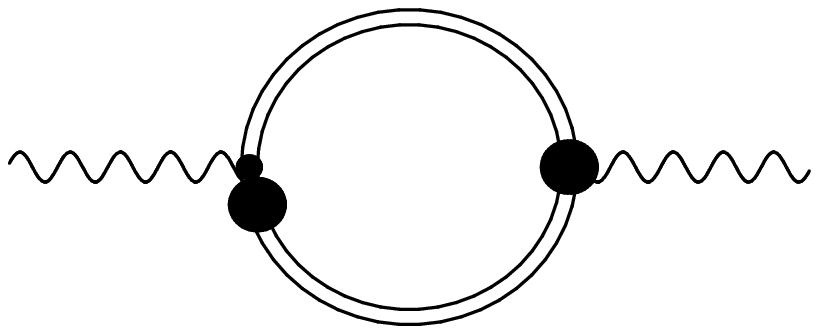}
\put(0.2,0.6){$=$}\hspace*{0.7cm}
\includegraphics[scale=0.41]{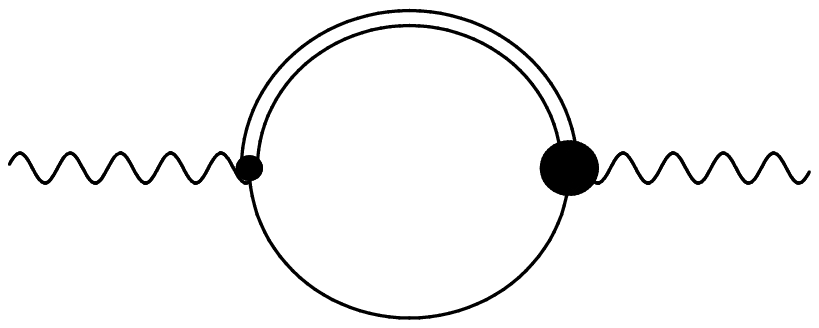}
\put(0.2,0.6){$+$}
\end{picture}
\nonumber\\
&&\qquad + e\,\frac{\delta}{\delta {\bf V}_y^b}
\,\mbox{tr}\Bigg[(T^a)^t \frac{m\bar D_x^2}{16\partial^2}
\Bigg(\frac{\delta^2\Gamma}{\delta\phi_z^+
\delta\tilde\phi_x^+}\Bigg)^{-1}_{z=z} - T^a \frac{m\bar
D_x^2}{16\partial^2}
\Bigg(\frac{\delta^2\Gamma}{\delta\tilde\phi_z^+
\delta\phi_x^+}\Bigg)^{-1}_{z=z} +\mbox{h.c.}\Bigg].
\end{eqnarray}

\noindent The single line in the second diagram corresponds to

\begin{equation}
- \frac{D^2 \bar D^2}{16\partial^2}\delta^8_{12}
\end{equation}

\noindent if fields at the ends 1 and 2 have opposite chirality,
and 0, if chirality of the ends is the same. Eq.
(\ref{Diagram_Identity}) can be easily verified using Feynman
rules and performing the differentiation with respect to the field
${\bf V}$.

We will calculate the expression

\begin{equation}\label{For_Calculation}
\frac{\partial}{\partial\ln\Lambda}\frac{\delta\Gamma}{\delta {\bf
V}^b_y \delta {\bf V}^a_x}\Bigg|_{p=0}.
\end{equation}

\noindent We note that the regularization by higher covariant
derivatives is essentially used here, because it allows
differentiating the integrand and taking the limit of zero
external momentum.

Using Eqs. (\ref{Inverse_Functions}), (\ref{Vertex1}), and
(\ref{Vertex2}) it is easy to see that

\begin{eqnarray}
&& e\,\frac{\partial}{\partial\ln\Lambda} \frac{\delta}{\delta
{\bf V}_y^b} \,\mbox{tr}\Bigg[(T^a)^t \frac{m\bar
D_x^2}{16\partial^2} \Bigg(\frac{\delta^2\Gamma}{\delta\phi_z^+
\delta\tilde\phi_x^+}\Bigg)^{-1}_{z=z} - T^a \frac{m\bar
D_x^2}{16\partial^2}
\Bigg(\frac{\delta^2\Gamma}{\delta\tilde\phi_z^+
\delta\phi_x^+}\Bigg)^{-1}_{z=z} +\mbox{h.c.} \Bigg]_{p=0}
=\nonumber\\
&& = e^2 C(R)\,\delta^{ab}\,\partial^2\Pi_{1/2}\delta^8_{xy}
\frac{\partial}{\partial\ln\Lambda} \int \frac{d^4q}{(2\pi)^4}
\frac{1}{q^2} \frac{d}{dq^2} \Bigg(\frac{m^2 J}{q^2 G^2 + m^2
J^2}\Bigg).\qquad
\end{eqnarray}

\noindent (All similar expressions are written in the Euclidean
space after the Weak rotation.)

\begin{figure}[h]
\hspace*{4.5cm}
\begin{picture}(0,0)
\put(0.1,-0.4){$y$} \put(2.7,-0.4){$z$} \put(1.6,-0.4){$x$}
\put(5,-0.4){$y$} \put(6.3,-0.4){$x$} \put(3.6,0.6){+}
\end{picture}
\includegraphics[scale=0.5]{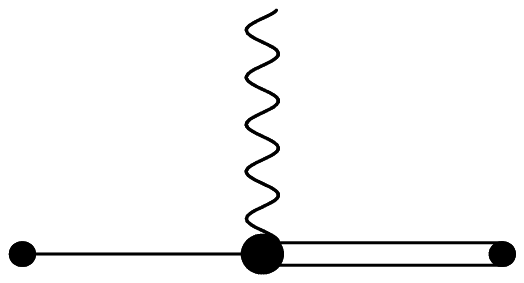}
\hspace*{2cm}
\includegraphics[scale=0.5]{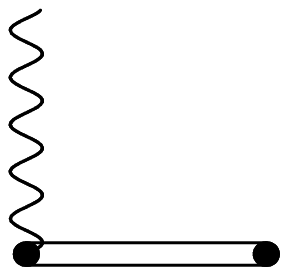}
\newline
\caption{The sum of subdiagrams in the Schwinger-Dyson equations.}
\label{Figure_Subdiagrams1}
\end{figure}

Let us proceed to the calculation of all other contributions in
Eq. (\ref{SD_Equation}). For this purpose we note that they
contain the sum of subdiagrams, presented in Fig.
\ref{Figure_Subdiagrams1}. The external line in these subdiagrams
corresponds to the background field ${\bf V}$, and expressions for
propagators were presented above. After simple transformations it
is easy to see that expressions for the sum of these subdiagrams
in the limit $p\to 0$ are written as

\begin{eqnarray}\label{Propagator1}
&& D_y^2 \bar D_y^2 \Big({\bf V}_{y} \frac{G D_y^2 \bar
D_y^2}{64\partial^2(\partial^2 G^2 + m^2 J^2)} \delta^8_{yz}\Big)
+ {\bf V}_{y} \frac{G D_y^2 \bar D_y^2}{4(\partial^2 G^2 + m^2
J^2)} \delta^8_{yz} = \nonumber\\
&& = - \int d^8x\,\frac{D_x^2}{16\partial^2} \delta^8_{xy} \Big( 2
\partial^2\Pi_{1/2} {\bf V}_{x} + i\bar D
\gamma^\mu\gamma_5 D {\bf V}_{x} \partial_\mu\Big) \frac{G \bar
D_z^2}{\partial^2 G^2 + m^2 J^2} \delta^8_{xz} -\nonumber\\
&& - D^a {\bf V}_{y} \frac{G D_a \bar D_z^2}{2(\partial^2 G^2 +
m^2 J^2)} \delta^8_{yz} - D^2 {\bf V}_{y} \frac{G \bar
D_z^2}{4(\partial^2 G^2 + m^2 J^2)} \delta^8_{yz}
\end{eqnarray}

\noindent or

\begin{eqnarray}\label{Propagator2}
&& = D_y^2 \bar D_y^2 \Big({\bf V}_{y} \frac{m J
D_y^2}{16\partial^2(\partial^2 G^2 + m^2 J^2)} \delta^8_{yz}\Big)
+ {\bf V}_{y} \frac{m J D_y^2}{\partial^2 G^2 + m^2 J^2}
\delta^8_{yz} = \nonumber\\
&& = \int d^8x\, \frac{D_x^2}{64\partial^2} \delta^8_{xy} \Big( 2
\partial^2\Pi_{1/2} {\bf V}_{x} + i\bar D \gamma^\mu\gamma_5 D
{\bf V}_{x} \partial_\mu \Big) \frac{m J \bar D_x^2
D_x^2}{\partial^2(\partial^2 G^2 +
m^2 J^2)}\delta^8_{xz} + \nonumber\\
&& + \frac{1}{8} D^a {\bf V}_{y} \frac{m J D_{ay} \bar D_y^2
D_y^2}{\partial^2(\partial^2 G^2 + m^2 J^2)} \delta^8_{yz} +
\frac{1}{16} D_y^2 {\bf V}_{y} \frac{m J \bar D_y^2
D_y^2}{\partial^2(\partial^2 G^2 + m^2 J^2)} \delta^8_{yz}
\end{eqnarray}

\noindent depending on the chirality of the ends. Terms in
subdiagrams, containing the integral over $d^8x$, have chiral
projection operators, acting on the ends $y$ and $z$. Due to
identities (\ref{Psi_Phi_Identites}), this allows using simpler
vertexes (\ref{Vertex1}) and (\ref{Vertex2}) for the calculations.
After simple, but rather long calculations we obtain that the
contribution of these terms to Eq. (\ref{For_Calculation}) is

\begin{eqnarray}
e^2 C(R)\,\delta^{ab}\,\partial^2 \Pi_{1/2} \delta^8_{xy}
\frac{\partial}{\partial\ln\Lambda} \int \frac{d^4q}{(2\pi)^4}
\frac{1}{q^2} \frac{d}{dq^2} \ln\Big(q^2 G^2 + m^2 J^2\Big).
\end{eqnarray}

\noindent It is also easy to see that a contribution of the terms,
which are proportional to $D^2 {\bf V}$ in Eqs.
(\ref{Propagator1}) and (\ref{Propagator2}), is 0. However, a
contribution of the terms, proportional to $D^a{\bf V}$ is
nontrivial and contains the unknown functions $f$ and $h$. It can
be written as

\begin{eqnarray}\label{Remaining_Diagrams}
\hspace*{-13.0cm}\begin{picture}(0,1.8)
\put(3.6,0.9){${\displaystyle =
\frac{\partial}{\partial\ln\Lambda}\,\mbox{tr} \int
d^8x\,d^8y\,{\bf V}_y D^a {\bf V}_x \frac{D_{az} \bar
D_z^2}{4i\partial^2} \frac{\delta^3\Gamma}{\delta j_z^+ \delta
{\bf V}_y \delta \phi_{0x}^+}\Bigg|_{z=x,p=0},}$}
\put(0.3,-0.5){$D^a V$} \put(3.0,-0.5){$V$} \put(2.6,1.3){$I_a$}
\hspace*{0.3cm}
\includegraphics[scale=0.4]{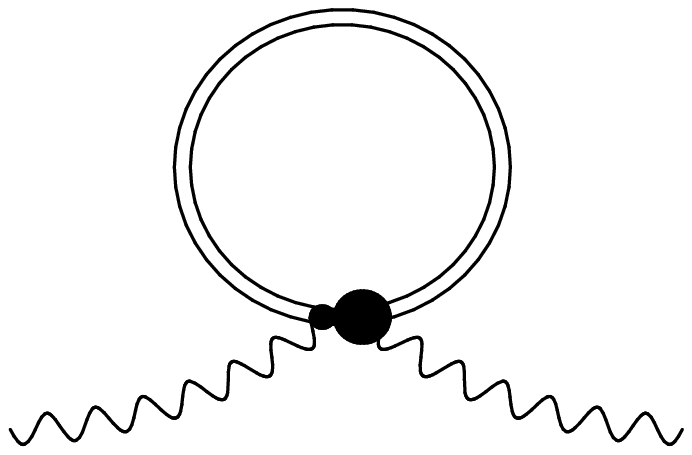}\hspace*{1cm}
\end{picture}
\end{eqnarray}

\noindent where the symbol $I_a$ (we use the notation in
\cite{Identity} here) means that in the expression for the exact
propagator it is necessary to make the substitution

\begin{equation}
\bar D^2 D^2 \delta^8_{xz} \to 2 \bar D^2 D^a \delta^8_{xz}.
\end{equation}

\noindent After substituting propagators and vertex functions the
considered contribution will be

\begin{equation}
e^2 C(R)\,\delta^{ab} \partial^2\Pi_{1/2} \delta^8_{xy}\,
\frac{\partial}{\partial\ln\Lambda} \int \frac{d^4q}{(2\pi)^4}
\Bigg(-\frac{m^2 J\,\Big(2 q^2 J' - (J-1)\Big)}{q^4\Big(q^2 G^2 +
m^2 J^2\Big)} - \frac{16 G f + 16 m^2 J h}{q^2 G^2 + m^2
J^2}\Bigg).
\end{equation}

Collecting all contributions to the effective action, we obtain
the following result ($Z=1$, a dependence on $Z$ will be restored
later):

\begin{eqnarray}\label{Result}
&& \frac{\partial}{\partial\ln\Lambda}\frac{\delta^2\Gamma}{\delta
{\bf V}_y^b \delta {\bf V}_x^a}\Bigg|_{p=0} = \ldots + e^2
C(R)\,\delta^{ab}
\partial^2\Pi_{1/2} \delta^8_{xy}\,
\frac{\partial}{\partial\ln\Lambda}\int\frac{d^4q}{(2\pi)^4}\Bigg\{
\frac{1}{q^2}\frac{d}{dq^2} \Bigg(\ln\Big(q^2 G^2
+\nonumber\\
&& + m^2 J^2\Big) + \frac{m^2 J}{q^2 G^2 + m^2 J^2}\Bigg)
-\frac{m^2 J\,\Big(2 q^2 J' - (J-1)\Big)}{q^4 \Big(q^2 G^2 + m^2
J^2\Big)} - \frac{16 G
f + 16 m^2 J h}{q^2 G^2 + m^2 J^2} -\nonumber\\
&& -\mbox{similar terms for the Pauli-Villars fields}\Bigg\},
\end{eqnarray}

\noindent where dots denote contributions of the gauge fields and
ghosts. Comparing the obtained result with the corresponding
expression for the supersymmetric electrodynamics, we see that
they differ in the multiplier $C(R)$. Nevertheless, one more
difference is possible: In the electrodynamics explicit
calculations show that there is an identity, which can be written
as equality to 0 of contribution (\ref{Remaining_Diagrams}). The
problem, if a similar identity is valid in the non-Abelian case,
is not yet solved. In general, it is not necessary, because there
are new diagrams, which contain a matter loop and vertexes with
self-action of the gauge superfield, in the non-Abelian case. That
is why in this paper we will not assume existence of such
identity.


\section{Exact Gell-Mann-Low function.}
\label{Section_Two_Point_Function} \hspace{\parindent}

Investigating the supersymmetric electrodynamics \cite{SD} reveals
that there are divergences only in the one-loop approximation if
the higher derivative regularization is used, while the
Gell-Mann-Low function coincides with the exact NSVZ
$\beta$-function. Let us consider a non-Abelian theory.

The Gell-Mann-Low function is defined by the dependence of finite
part of the two-point Green function on the momentum in the limit
$m\to 0$. And so, let us consider the massless case and write the
renormalized two-point Green function as

\begin{equation}
\Gamma^{(2)}_V = - \frac{1}{16\pi} \mbox{tr}\int
\frac{d^4p}{(2\pi)^4}\,{\bf V}(-p)\,\partial^2\Pi_{1/2} {\bf
V}(p)\, d^{-1}(\alpha,\mu/p),
\end{equation}

\noindent where $\alpha$ is a renormalized coupling constant. It
is known \cite{Bogolyubov,Itzykson} that the function $d$ is
invariant under renormgroup transformations, i.e. does not depend
on a choice of the renormalization prescription. The Gell-Mann-Low
function, which we will denote by $\beta(\alpha)$, is given by

\begin{equation}\label{Gell-Mann-Low_Definition}
\beta\Big(d(\alpha,\mu/p)\Big) = - \frac{\partial}{\partial \ln p}
d(\alpha,\mu/p).
\end{equation}

\noindent Because the function $d$ is renorminvariant, the
function $\beta(\alpha)$ will not also depend on a particular
renormalization prescription. This is its difference from the
function $b(\alpha)$, which is constructed from the renormalized
coupling constant by the differentiation with respect to the
logarithm of the normalization point.

In order to obtain physical results it is necessary to calculate
the function $d$ in the massive case, and then to impose the
boundary condition

\begin{equation}
d(\alpha,\mu/p,m/\mu)\Big|_{p=0} = \alpha_{ph},
\end{equation}

\noindent where $\alpha_{ph}$ is an experimentally measured
coupling constant, at a given $\mu$. Moreover, it is necessary to
impose one more boundary condition to the Green functions of the
matter superfield:

\begin{equation}
G(\alpha,\mu/p,m/\mu)\Big|_{p^2=m^2_{ph}} = \frac{m}{m_{ph}}
J(\alpha,\mu/p,m/\mu)\Big|_{p^2=m^2_{ph}},
\end{equation}

\noindent in which $m_{ph}$ denotes a physical mass. (This
equality assures that the propagator of the matter superfield will
have a pole at the physical mass.) The boundary conditions written
above allow expressing the parameters $\alpha$ and $m$ (at a given
$\mu$) through the physically observed values. Nevertheless, the
results of this paper do not allow finding explicit form of the
functions $d$, $G$, and $J$ in the massive case. And so, we can
not relate $\alpha$ and $m$ with the experimentally observed
values. However, it is possible to obtain an expression for the
contribution of matter superfields to the Gell-Mann-Low function:

If $d_0$ denotes the function $d$ calculated at $Z=1$, the
following equation takes place in accordance with Eq.
(\ref{Result})

\begin{eqnarray}\label{Effective_Action}
&& - \frac{\partial}{\partial\ln p}\,
d_0^{-1}(\alpha_0,\Lambda/p)\Bigg|_{p=0} =
\frac{\partial}{\partial\ln \Lambda}\,
d_0^{-1}(\alpha_0,\Lambda/p)\Bigg|_{p=0} = -16\pi C(R)\,
\frac{\partial}{\partial\ln\Lambda} \times\nonumber\\
&& \times \int\frac{d^4q}{(2\pi)^4} \frac{1}{2q^2}\frac{d}{dq^2}
\Bigg\{\ln(q^2 G^2) + X(q^2/\Lambda^2) - \sum\limits_i c_i
\ln\Big(q^2 G_{PV}^2 + M_i^2 J_{PV}^2\Big)
-\nonumber\\
&& -\sum\limits_i c_i \frac{M_i^2 J_{PV}}{q^2 G_{PV}^2 + M_i^2
J_{PV}^2} - \sum\limits_i c_i X(q^2/\Lambda^2,a_i)\Bigg\},
\end{eqnarray}

\noindent where the dimensionless functions $X$ are obtained from
the equations

\begin{eqnarray}\label{X_Definition}
&& \frac{\partial X(q^2/\Lambda^2)}{\partial q^2}
= - \frac{16f}{G};\nonumber\\
&& \frac{\partial X(q^2/\Lambda^2,a_i)}{\partial q^2} = -
\frac{M_i^2 J\,\Big(2 q^2 J' - (J-1)\Big)}{q^4\Big(q^2 G^2 + M_i^2
J^2\Big)} - \frac{16 G f + 16 M_i^2 J h}{q^2 G^2 + M_i^2 J^2},
\end{eqnarray}

\noindent in which $M_i = a_i\Lambda$.

The obtained integral is reduced to the total derivative in the
four-dimensional spherical coordinates, only the substitution at
the low limit being different from 0 \cite{SD}:

\begin{eqnarray}\label{Equation_For_D0}
&& \frac{\partial}{\partial\ln \Lambda}\,d_0^{-1}\Bigg|_{p=0} =
\frac{C(R)}{2\pi}\,\frac{\partial}{\partial\ln\Lambda} \Bigg\{\ln
G(0)^2 + X(0) - \sum\limits_i
c_i \ln\Big(M_i^2 J_{PV}(0,a_i)^2\Big) -\nonumber\\
&& -\sum\limits_i c_i \frac{1}{J_{PV}(0,a_i)} - \sum\limits_i c_i
X(0,a_i)\Bigg\} = - \frac{1}{\pi} C(R) \Bigg(1-\frac{\partial\ln
G}{\partial\ln \Lambda} - \frac{1}{2} \frac{\partial
X}{\partial\ln\Lambda}\Bigg)\Bigg|_{q=0}
\end{eqnarray}

\noindent Here we took into account that the functions
$X(q^2/\Lambda^2,a_i)$ and $J_{PV}(q^2/\Lambda^2,a_i)$ in the
limit $q\to 0$ tended to some finite constants, because they were
defined by convergent (even in the limit $\Lambda\to\infty$),
dimensionless integrals, which did not contain infrared
divergences. Because both parts of Eq. (\ref{Equation_For_D0})
depend on $\alpha_0$ and $\Lambda/p$, it allows finding the
expression for $d_0^{-1}$ up to an insignificant numerical
constant

\begin{equation}
d_0^{-1}(\alpha_0,\Lambda/p) = - \frac{1}{\pi} C(R) \Big(\ln
\frac{\Lambda}{p}- \ln G(\alpha_0,\Lambda/p) - \frac{1}{2}
X(\alpha_0,\Lambda/p)\Big) + \mbox{const}.
\end{equation}

Making substitution (\ref{M_Substitution}) in Eq.
(\ref{Equation_For_D0}) it is possible to restore the dependence
of the effective action on the renormalization constant $Z$:

\begin{equation}
d^{-1}(\alpha,\mu/p) = d_0^{-1}(\alpha_0,\Lambda/p) +
\frac{C(R)}{\pi}\ln Z(\alpha,\Lambda/\mu).
\end{equation}

\noindent Differentiating this expression with respect to the
momentum $p$, we obtain the Gell-Mann-Low function:

\begin{eqnarray}\label{Massless_D}
&& - \frac{\partial}{\partial\ln p}\,d^{-1} = - \frac{1}{\pi} C(R)
\Bigg(1+\frac{\partial\ln G}{\partial\ln p} + \frac{1}{2}
\frac{\partial X}{\partial\ln p}\Bigg).
\end{eqnarray}

\noindent Using Eq. (\ref{X_Definition}), we rewrite this
contribution of the matter superfield as

\begin{equation}\label{Matter_Contribution}
- \frac{C(R)}{\pi}\Big(1-\gamma(\alpha) - \lim\limits_{p\to
0}\frac{16 p^2 f}{G}\Big),
\end{equation}

\noindent where $\gamma(\alpha)$ is the anomalous dimension of the
matter superfield:

\begin{equation}
\gamma\Big(d(\alpha,\mu/p)\Big) = -\frac{\partial}{\partial\ln p}
\ln ZG(\alpha,\mu/p).
\end{equation}

\noindent (Expression (\ref{Matter_Contribution}) contributes to
the Gell-Mann-Low function and should be compared with the
numerator of Eq. (\ref{NSVZ_Beta}).) Let us again mention that the
obtained result does not depend on a choice of renormalization
scheme.

We note that the contribution of the last term in Eq.
(\ref{Matter_Contribution}) was always 0 in all diagrams
calculated up to now. Moreover, if the identity

\begin{equation}\label{New_Identity}
\mbox{tr}\,\frac{\partial}{\partial\ln\Lambda}\int
d^8x\,d^8y\,{\bf V}_y D^a {\bf V}_x \frac{D_{az} \bar
D_z^2}{4i\partial^2} \frac{\delta^3\Gamma}{\delta j_z^+ \delta
{\bf V}_y \delta \phi_{0x}^+}\Bigg|_{z=x, p=0} = 0,
\end{equation}

\noindent is valid in the non-Abelian case, this will be always
true. Nevertheless, in the non-Abelian case this statement
requires an additional verification. That is why we for the
generality write the correction, corresponding to a nontrivial
contribution of the function $f$.

We note that the function $f$ can actually affect only diagrams in
three and more loops. Therefore, if such contribution appeared
calculating the scheme dependent function $b(\alpha)$, it could be
always eliminated by a proper redefinition of the coupling
constant.

Finally we note that due to the similarity of the result with the
electrodynamics case, if the contribution of the function $f$ is
absent, there are divergences only in the one-loop approximation.
Details of the corresponding calculation are presented in Ref.
\cite{SD}. This completely agrees with the structure of the
anomalies supermultiplet and is a particular feature of higher
covariant derivative regularization using \cite{HD_And_DRED}.


\section{Conclusion}
\label{Section_Conclusion}
\hspace{\parindent}

In this paper a contribution of the matter superfields to the
two-point Green function of the gauge field is calculated by the
Schwinger-Dyson equations and Slavnov-Taylor identities for the
supersymmetric Yang-Mills theory regularized by higher covariant
derivatives. This regularization is very essential, because the
method of calculating, used here, is based on the differentiation
with respect to the regularization parameter and taking the limit
of zero momentum.

In many respects the obtained results are similar to the
corresponding results in the electrodynamics. However, it is
necessary to mention some differences:

1. The calculation was made using the background field method,
because this method allows preserving the manifest gauge
invariance of the effective action. Without its using it would be
also necessary to calculate the renormalization constant for the
gauge superfield that make calculations essentially more
complicated.

2. There is the multiplier $C(R)$, defined by Eq. (\ref{C(R)}) in
the non-Abelian case. The same multiplier is also present in the
exact NSVZ $\beta$-function.

3. Deviations of the result from the exact $\beta$-function, due
to the contribution of the function $f$, which can not be
determined from the Slavnov-Taylor identity, are possible
hypothetically. Such contributions seem to disappear in the
electrodynamics due to a new identity for the Green functions.
However, there are diagrams, containing vertexes with self-action
of the gauge field, in the non-Abelian case, which can break the
additional identity. Contributions of these diagrams were not yet
calculated explicitly in the considered regularization. That is
why we also present the result for the $\beta$-function, if there
is a nontrivial contribution of the function $f$.

We note that according to the above discussion, it is possible to
point two possible reasons why the three-loop calculations with
the dimensional reduction differ from the NSVZ $\beta$-function:

1. There is a nontrivial contribution of the function $f$.

2. In preceding papers instead of the Gell-Mann-Low function, the
$\beta$-function, defined as a derivative of the renormalized
coupling constant in the $\overline{\mbox{MS}}$-scheme, is
calculated.

At present we can not affirm which of this variants (or both ones
simultaneously) is realized. This requires either explicit
calculating or finding an origin of identity (\ref{New_Identity}).

Moreover, it is necessary to calculate a contribution of the gauge
fields and ghosts using Schwinger-Dyson equations and
Slavnov-Taylor identities. Now this work is in progress.

\bigskip
\bigskip

\noindent {\Large\bf Acknowledgments.}

\bigskip

\noindent This paper was supported by the Russian Foundation for
Basic Research (Grant No. 05-01-00541).


\end{document}